\documentclass[12pt]{article}
\usepackage[utf8x]{inputenc}
\usepackage[active]{srcltx} 
\usepackage{epsfig}
\usepackage{multirow}
\usepackage{graphicx}
\usepackage{subfigure}
\usepackage[singlespacing]{setspace}
\usepackage[small,bf]{caption}
\title{Emission spectra of \emph{p-Si} and \emph{p-Si:H} models generated by \emph{ab initio} molecular dynamics methods}
\author{E. R. L. Loustau$^{ac}$, Ariel A. Valladares$^{b}$\\
$^{a}$ Centro de Investigaci\'on en Energ\'ia, \\ Universidad Nacional Aut\'onoma de M\'exico (UNAM)\\ A. P.34, CP.62580 Temixco, Morelos, M\'exico.\\
$^{b}$ Instituto de Investigaciones en Materiales (UNAM)\\ A. P.72, CP.04510 D.F., M\'exico.\\
$^{c}$ Centro de Ciencias de la Complejidad (UNAM)\\ A. P.70472, CP.04510, D.F., M\'exico.\\}

\begin{document}

\maketitle

\begin{abstract}
We created 4 \emph{p-Si} models and 4 \emph{p-Si:H} models all with 50\% porosity. The models contain 32, 108, 256 and 500 silicon atoms with a pore parallel to one of the simulational cell axes and a regular cross-section. We obtained the densities of states of our models by means of ab initio computational methods. We wrote a code to simulate the emission spectra of our structures considering particular excitations an decay conditions. After comparing the simulated spectra with the experimental results, we observe that the position of the maximum of the emission spectra might be related with the size of the silicon backbone for the \emph{p-Si} models as the quantum confinement models say and with the hydrogen concentration for the \emph{p-Si:H} structures. We conclude that the quantum confinement model can be used to explain the emission of the \emph{p-Si} structures but, in the case of the \emph{p-Si:H} models it is necessary to consider others theories.
\end{abstract}

\section{Introduction}

Since the discovery of the \emph{p-Si} photoluminiscence (PL) at room temperature realized by L.T. Canham in 1990 \cite{Canham}, a lot of studies have been carried out to understand the origin of this property. The possibility that \emph{p-Si} could be applied to photonic and optoelectronic devices, motivated the study of the dependency of its PL to the temperature, chemical attack variables and molecules on its surface. Despite the big amount of scientific reports about the \emph{p-Si} PL, until now, we have not a model that reproduce the characteristics of all experimental PL spectra reported in the bibliography.\\

During the research of the origins of the \emph{p-Si} PL, some models about the nature of the PL were proposed; Cullis \emph{et al.}\cite{Cullis} say that all these models could be reorganized in 6 mayor groups. We only review the fundamentals of the confinement quantum model (QC), and of the hydrogen-groups model (HS) because we will use them to interpret our results.\\

The QC model proposed by Canham in 1990 was the first model created with the aim to explain the efficient PL of the \emph{p-Si} at room temperature. In this model, because of the effects of the quantum confinement, the energy gap of the silicon nanostructures increase, and their photoluminescence appear in the visible range of the electromagnetic spectrum. In accord to the QC model, if the size of the silicon nanostructures (where the recombination process take place) decrease, then the energy of the photons created increase. The QC model is the most successful theory about the nature of the \emph{p-Si} photoluminescence, because it explain various experimental observations, with an exception: The PL spectra of oxidized \emph{p-Si} structures.\\

In the other hand, the HS model consider that the hydrogen atoms on the \emph{p-Si} surface (which satisfy the dangling bonds), could be involved in the origin of PL of this material. The HS model was proposed after observing that if the hydrogen atoms are removed from the \emph{p-Si} surface, by a thermic procedure, then the intensity of the PL decrease dramatically (but still exist because of the QC model argument). Since the \emph{p-Si} with a rusty surface present an intense PL, we understand that, the porous silicon with an hydrogen passivated surface (\emph{p-Si:H}) is just one of the luminescent forms of the \emph{p-Si} (Cullis \emph{et al.}).\\

Photoluminescence is a difficult phenomenon to simulate, since the dynamics of the excitation and decay of the electrons require precise calculations of the excited states and decaying mechanisms. Also, metastable states may take part in the process with short, but relevant, mean free lives that require time-dependent quantum mechanical calculations \cite{Torchynska}, \cite{Calcott}. Instead, we have decided to initially investigate the photoemission spectrum as a crude approximation to the photoluminescence phenomenon.\\

Bearing in mind the QC and HS models, we propose a simple photoemission model in which the excited electrons of the conduction band, decay successively into the unoccupied electronic states of the valence band (see figure \ref{Modeca}). The emission spectra that we obtain for the \emph{p-Si} and \emph{p-Si:H} structures are comparable to the experimental spectra reported by Cullis \emph{et al}.\\

We are aware that our simulated \emph{p-Si} structures do not have experimental counterpart, because the number of silicon atoms involved in each simulation is finite due to the computational cost of the ab initio calculations. Also, in our simulations is not possible to reproduce all the chemical molecules that generate after the chemical attack of the \emph{p-Si} surface, because depending of the atmosphere where the attack was realized, molecules or atoms besides hydrogen atoms, could satisfy the dangling bonds of the \emph{p-Si} surface modifying the form of its PL spectrum.\\

However, until now the doubts about the origins of the \emph{p-Si} PL remains, so, our work contributes to calculate the photoemission spectra of \emph{p-Si} and \emph{p-Si:H} structures with ab initio molecular dynamics methods. In the following section, we describe the construction method used to generate the \emph{p-Si} and \emph{p-Si:H} structures, and the way we obtain their photoemission spectra.\\

\section{Method}

The \emph{p-Si} structures were created using the \emph{Cerius2} interface \cite{Cerius}. We reproduced the silicon cell diamond-like which contains 8 silicon atoms 2 times along each of the coordinate axes to obtain a 64 silicon-atom supercell, 3 times to increase the number of atoms to 216, 4 times for a supercell with 512 atoms and finally, 5 times along each direction to obtain 1000 atoms in the silicon supercell.\\

Once the supercells were created, we eliminated 50\% of their silicon atoms carving a central pore parallel to the z-axis and with a regular cross section. After eliminating 50\% of the silicon atoms contained in each supercell, we obtained 4 \emph{p-Si} structures with 32, 108, 256 and 500 silicon atoms. The pore of the \emph{p-Si} structure with 32 silicon atoms has an irregular transversal section because the great number of atoms removed from the supercell. In the other hand, the pore of the supercell with 500 silicon atoms shows a regular cross section because even if we remove 50\% of their atoms, its silicon backbone is bigger than the one for the 32 silicon atoms model.\\

\begin{figure}
\begin{center}
\subfigure[pSi32]{\label{psi32}\includegraphics[width=0.4\textwidth, height=4.0 cm]{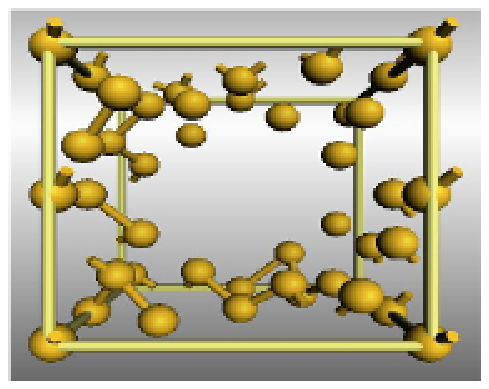}}
\subfigure[pSi108]{\label{pSi108}\includegraphics[width=0.4\textwidth, height=4.0 cm]{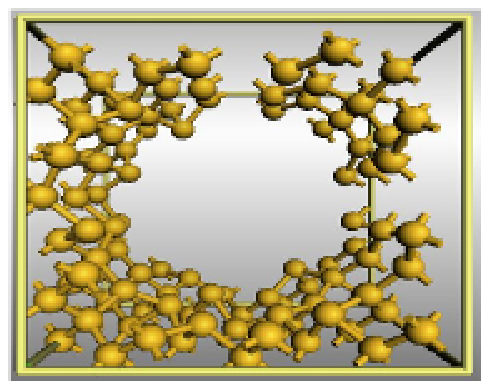}}
\subfigure[pSi256]{\label{pSi256}\includegraphics[width=0.4\textwidth, height=4.0 cm]{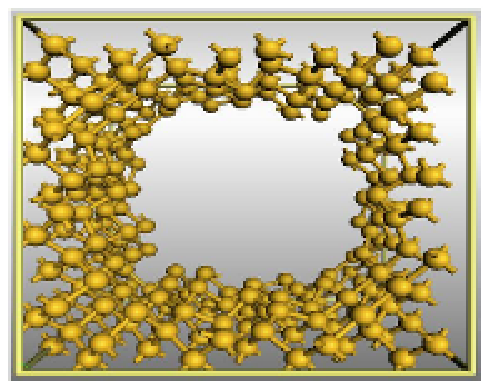}}
\subfigure[pSi500]{\label{pSi500}\includegraphics[width=0.4\textwidth, height=4.0 cm]{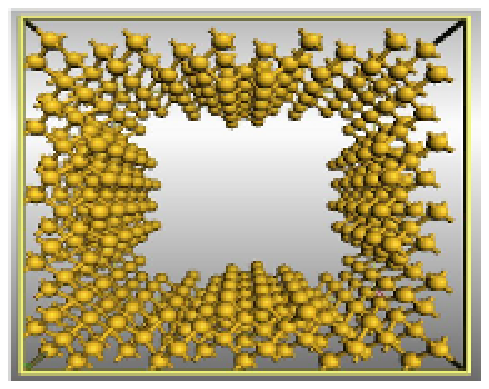}}
\caption{\label{ModxpSi}. \emph{p-Si} optimized models with 32 (a), 108 (b), 256 (c) y 500 (d) silicon atoms. All the models have 50 \% porosity.}
\end{center}
\end{figure}

The 4 structures of \emph{p-Si} were subjected to a geometry optimization process with the objective of finding their equilibrium configurations (see figures \ref{psi32} to \ref{pSi500}). We perform the geometry optimization process with \emph{FAST STRUCTURE} a code included in the \emph{CERIUS2} interface based in the Lin and Harris approach \cite{Lin}, \cite{Harris}, and the VWN \cite{vosko} functional of correlation energy.\\

In table \ref{uno} we indicate the computational parameters used for the geometry optimizations of the 4 \emph{p-Si} structures. The integration grid refers to the space were the atomic functions, atomic densities, Coulomb potential, exchange and correlation potentials are represented. To calculate the hamiltonian matrix elements it is necessary to put each atom in the center of the integration grid. A coarse integration grid leads to a poor representation of the atomic functions, whereas a fine grid is the best approximation but its computational cost is elevated. The minimal bases indicate that we use the valence atomic orbital to construct the wave function, and the minimal density indicates that the electronic density should be obtained just using the valence atomic orbital as well. Finally, the cut-off parameter corresponds to the distance at which the wave function vanish.\\

\begin{figure}
\begin{center}
\subfigure[FDRpSi32]{\label{FDR32}\includegraphics[width=0.4\textwidth, height=4.0 cm]{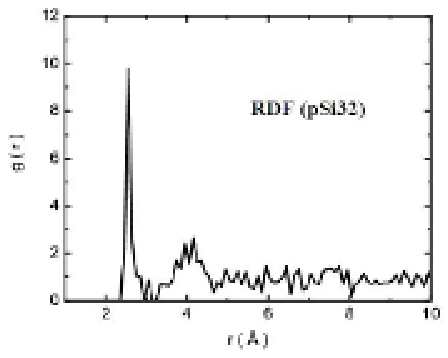}}
\subfigure[FDRpSi108]{\label{FDR108}\includegraphics[width=0.4\textwidth, height=4.0 cm]{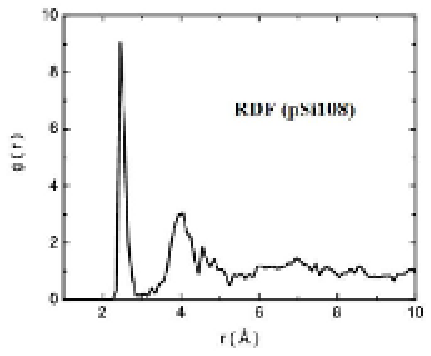}}
\subfigure[FDRpSi256]{\label{FDR256}\includegraphics[width=0.4\textwidth, height=4.0 cm]{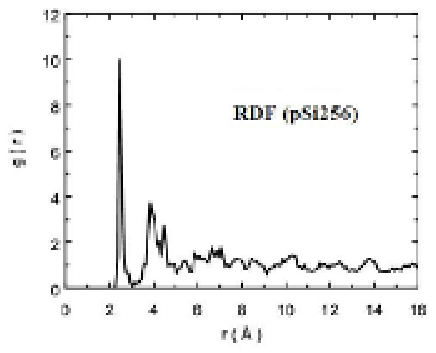}}
\subfigure[FDRpSi500]{\label{FDR500}\includegraphics[width=0.4\textwidth, height=4.0 cm]{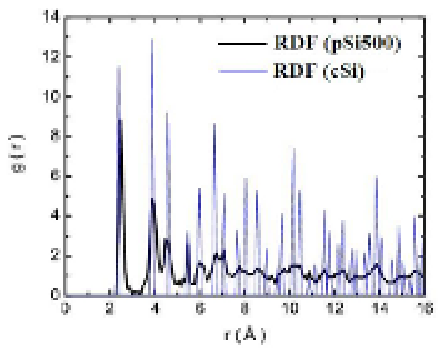}}
\caption{Radial distribution functions of the \emph{p-Si} optimized models with 32 (a), 108 (b), 256 (c) and 500 (d) silicon atoms.}
\end{center}
\end{figure}

\begin{figure}
\begin{center}
\includegraphics[width=0.9\textwidth, height=7.0 cm]{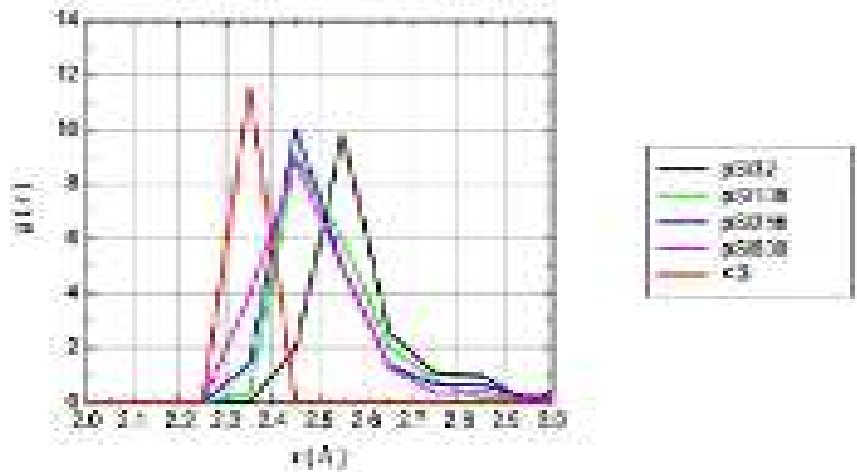}
\caption{\label{secuencia}. Interatomic distance variation of the \emph{p-Si} optimized models, as a function of the size of their silicon backbones.}
\end{center}
\end{figure}

\begin{figure}
\begin{center}
\includegraphics[width=0.7\textwidth, height=7.0 cm]{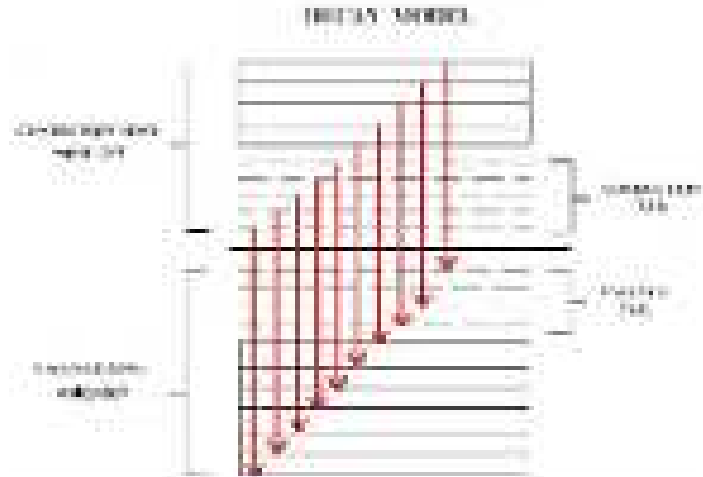}
\caption{\label{Modeca}. Order (considered in the EDOxEDO.f code.), of the excited electrons decay from the conduction band to the valence band.}
\end{center}
\end{figure}

At the end of the geometry optimizations of our \emph{p-Si} structures, the interface brings forth their radial distribution functions (figures \ref{FDR32} to \ref{FDR500} and \ref{secuencia}) and densities of states. We used part of the values of the densities of states of the \emph{p-Si} models as the input information for our emission code (EDOxEDO.f). We consider that the electronic states of the conduction density are unoccupied and those of the valence density are occupied. Following the decay order proposed by our code (figure \ref{Modeca}), the excited electrons of the conduction states decay into the valence density states emitting a photon, whose energy is equal to the energetic difference between the two electronic states involved. The energies of the photons are computed as an histogram and then, we adjust a gaussian curve to the histogram to compare it with the \emph{p-Si} PL experimental spectrum reported by Cullis \emph{et al.} (Cullis \emph{et al.}). To calculate the emission of all our structures, we only consider the electronic states that belong to a 2 eV interval in the valence and conduction densities of states. We decide to work out with a 2 eV interval states because experimentally, the majority of the PL excitation sources have energies of this order of magnitude.\\

\begin{figure}
\begin{center}
\subfigure[pSi32H42]{\label{pSi32H42}\includegraphics[width=0.4\textwidth, height=4.0 cm]{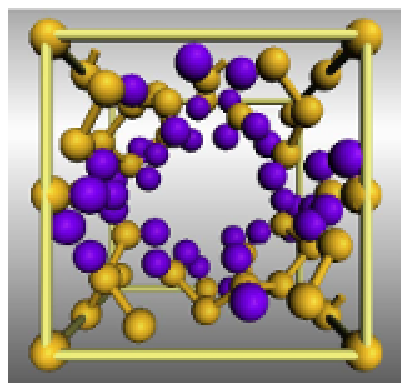}}
\subfigure[pSi108H68]{\label{pSi108H68}\includegraphics[width=0.4\textwidth, height=4.0 cm]{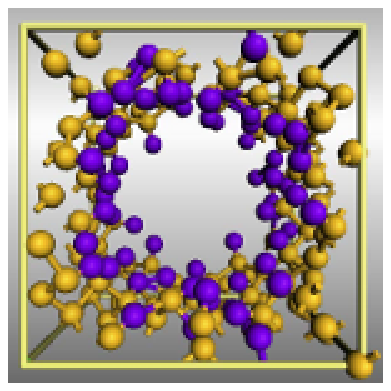}}
\subfigure[pSi256H153]{\label{pSi256H153}\includegraphics[width=0.4\textwidth, height=4.0 cm]{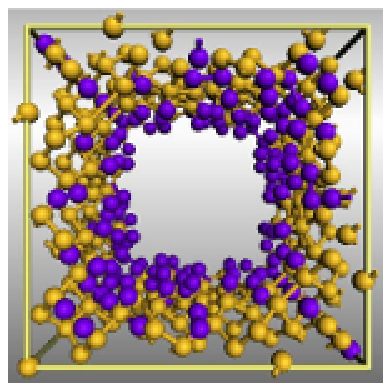}}
\subfigure[pSi500H196]{\label{pSi500H196}\includegraphics[width=0.4\textwidth, height=4.0 cm]{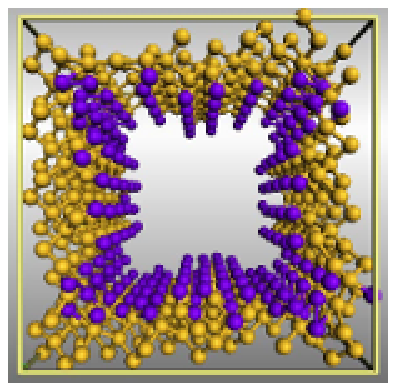}}
\caption{\label{ModxpSiH}. \emph{p-Si:H} optimized structures with 32 silicon atoms and 42 of hydrogen (a), 108 silicon atoms and 68 of hydrogen (b), 256 silicon atoms and 153 of hydrogen (c) and 500 silicon atoms and 196 of hydrogen (d).}
\end{center}
\end{figure}

The \emph{p-Si:H} models were constructed taking the \emph{p-Si} optimized structures as the initial atomic arrangement. We saturated with hydrogen the \emph{p-Si} optimized structures. After the saturation with hydrogen, we applied a geometry optimization to the \emph{p-Si:H} structures using the parameters of table \ref{uno}. In figures \ref{pSi32H42} to \ref{pSi500H196} we present the 4 optimized structures of \emph{p-Si:H}. We obtain the valence and conduction states from the densities of states of the \emph{p-Si:H} models in the same way than for the \emph{p-Si} structures. Again, to obtain the emission spectra of the \emph{p-Si:H} models we just consider the electronic states, in the valence and conduction densities, that are in an 2 eV interval. The order of decay of the excited electronic states into the unoccupied valence states is shown in figure \ref{Modeca}. The distribution of the photon energies is obtained after a gaussian fit.\\

In the results section we exhibit the photoemission spectra of the \emph{p-Si} and \emph{p-Si:H} models. We compare our photoemission spectra with the PL spectrum reported by Cullis \emph{et al.}.

\begin{table}
\begin{center}
\begin{tabular}{|c|c|}
  \hline
  \small{Parameter} & \small{Option}\\
  \hline
  \small{Integration grid} & \small{Coarse}\\
  \hline
  \small{Bases}&\small{Minimal}\\
  \hline
  \small{Density}&\small{Minimal}\\
  \hline
  \small{Cut off radius}&\small{5 \AA}\\
  \hline
\end{tabular}
\caption{\label{uno}. Computational parameters used for the geometry optimization process implemented by the \emph{FAST STRUCTURE} code for the \emph{p-Si} and \emph{p-Si:H} models.}
\end{center}
\end{table}

\section{Results and discussion}

As we can see from figures \ref{FDR32} to \ref{FDR500}, the radial distribution functions of the \emph{p-Si} structures become similar to the radial distribution function of a crystalline silicon structure, until the number of silicon atoms reach 500 (figure \ref{FDR500}). The variation of the interatomic distance of the \emph{p-Si} structures as a function of their silicon backbones size is shown in figure \ref{secuencia}. The backbone of the \emph{p-Si} structure with 500 silicon atoms is thicker than the one of the model with just 32 silicon atoms, therefore, the thinner silicon backbone is really deformed after a geometry optimization process because of the presence of the porous in the simulation cell, then the most probable distance between silicon neighbors increase to 2.55 {\AA}. In the other hand, the backbone of the \emph{p-Si} structure with 500 silicon atoms, is thick enough to remain as a crystalline silicon structure despite the central porous, so then the distance between first neighbors is 2.35 {\AA}, as for a crystalline silicon structure.\\

\begin{figure}
\begin{center}
\subfigure[pSi32]{\label{Ajus32}\includegraphics[width=0.4\textwidth, height=4.0 cm]{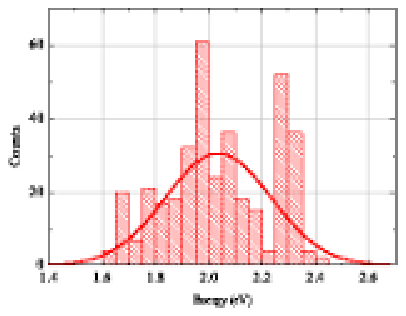}}
\subfigure[pSi108]{\label{Ajus108}\includegraphics[width=0.4\textwidth, height=4.0 cm]{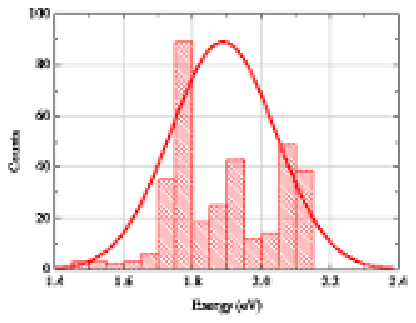}}
\subfigure[pSi256]{\label{Ajus256}\includegraphics[width=0.4\textwidth, height=4.0 cm]{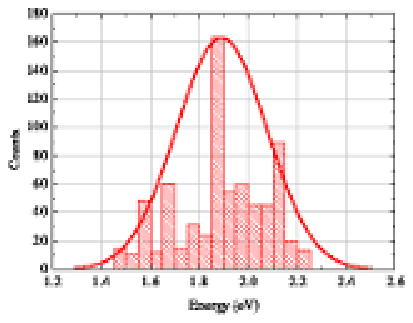}}
\subfigure[pSi500]{\label{Ajus500}\includegraphics[width=0.4\textwidth, height=4.0 cm]{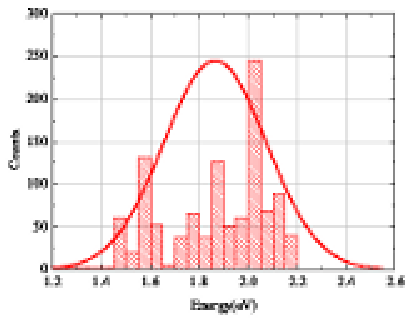}}
\caption{Gaussian fitting of the \emph{p-Si} model histograms (a) with 32, (b) 108, (c) 256, and (d) 500 silicon atoms.}
\end{center}
\end{figure}

\begin{center}
\begin{figure}
\subfigure[Emission spectra of the \emph{p-Si} models]{\label{PLpSi}\includegraphics[width=0.49\textwidth, height=5.0 cm]{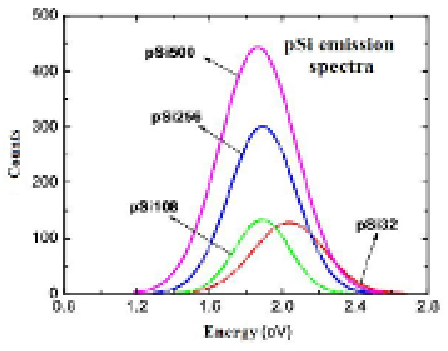}}
\subfigure[Simulation and experiment]{\label{CullisPL}\includegraphics[width=0.49\textwidth, height=5.0 cm]{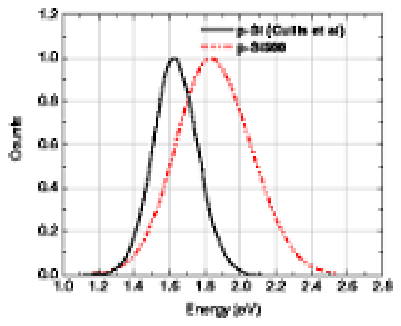}}
\caption{(a) Emission spectra of the 4 \emph{p-Si} structures calculated with the EDOxEDO.f code. (b) \emph{p-Si} PL spectrum reported by Cullis \emph{et al.} (black curve) and the emission spectrum of our \emph{p-Si} model with 500 silicon atoms (red curve).}
\end{figure}
\end{center}

Figures \ref{Ajus32}, \ref{Ajus108}, \ref{Ajus256} and \ref{Ajus500} show the gaussian curves fitted to the histogram of each \emph{p-Si} model obtained by our EDOxEDO.f code. In figure \ref{PLpSi} we present the emission spectra (just the gaussian curves) of the 4 \emph{p-Si} optimized structures; for these 4 structures their photoemission energy range from 1.2 eV to 2.8 eV, its mean maximum photoemission energy is 1.94 eV, and as we can observe, the intensity of the emission increase if the size of the backbone increase, but its maximum energy decrease in accord to the results of Cullis \emph{et al.}.\\

In figure \ref{CullisPL} we compare the \emph{p-Si} PL spectrum reported by Cullis \emph{et al.} and the emission spectrum of our \emph{p-Si} model with 500 silicon atoms; from this figure we can appreciate that our spectrum is larger than the Cullis \emph{et al.} but both have a similar form. Also, the maximum photoemission energy of our spectrum is 0.34 eV to the blue of the electromagnetic spectrum considering the va

\begin{figure}
\begin{center}
\includegraphics[width=0.49\textwidth, height=5.0 cm]{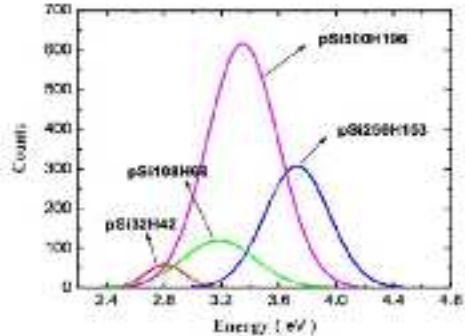}
\caption{\label{PLpSiHmod}. Emission spectra of the four \emph{p-Si:H} models calculated with de EDOxEDO.f code.}
\end{center}
\end{figure}

In figure \ref{PLpSiHmod} we superpose the emission spectra of the 4 models of \emph{p-Si:H} calculated by the EDOxEDO.f code. We did not find experimental photoemission spectra for the \emph{p-Si:H} to compare to our results.\\

\section{Conclusions}

We observed that the form of the emission spectra, obtained after a gaussian fitting of the histograms calculated by the EDOxEDO.f code, is similar to the form of the Cullis \emph{et al.} spectrum. The maximum width of our spectra is 1.6 eV, their minimal energy is 1.2 eV and the maximal one is 2.8 eV. The mean emission energy calculated was 1.94 eV, this is, 0.34 eV toward the blue of the electromagnetic spectrum, with respect the Cullis \emph{et al.} maximum energy.\\

Is clear from the figure \ref{PLpSi} that decreasing the number of silicon atoms in the simulation cell, moves toward bigger energies the maximal energy of the emission spectra. We think that this fact is due to the size of the quantum confinement region in the silicon backbone; for example, in the 32 \emph{p-Si} structure, the silicon backbone is smaller than the one of the 500 silicon atoms model and, in accord with the quantum confinement model, his emission energy should be bigger than the one of the \emph{p-Si} 500 atoms model.\\

From figure \ref{PLpSi} we conclude that decreasing the number of silicon atoms in the \emph{p-Si} supercells results in a smaller emission intensity. We think that because the \emph{p-Si} 500 silicon atoms model has a crystalline structure (\ref{FDR500}) compared with the 32 atoms structure (\ref{FDR32}), the radiative recombination of the charges dominates in the crystalline structures of the \emph{p-Si} and is not significant in the amorphous ones.\\

We observe from figure \ref{CullisPL} that the emission spectra of our \emph{p-Si} 500 silicon atoms model is toward the right of the electromagnetic spectrum, with respect to the PL maximum reported by Cullis \emph{et al.}. We consider that the size of the supercell is the direct cause and probably, if we could be able to increase the number of atoms, to thousand of they to construct the \emph{p-Si} models, theirs emission maximus will be near the experimental ones reported.\\

\section{Acknowledgements}
 Emilye Rosas Landa Loustau acknowledges the financial support of CONACyT during his PhD studies; she also wants to thank  Dr. Lorenzo Pavesi, and Dr. Juan Carlos Alonso Huitr\'on because of their fruitful discussions.

\end{document}